\documentclass[amsmath,superscriptaddress,showpacs,aps,twocolumn,pra]{revtex4-1}
\usepackage{graphicx}
\usepackage{rotating}
\usepackage{amssymb}
\usepackage{mathptmx}
\usepackage[usenames,dvipsnames]{color}
\usepackage[colorlinks=true,linkcolor=blue,citecolor=BrickRed]{hyperref}

\begin{document}
\title{Mott transition in a two-leg Bose-Hubbard ladder under an artificial magnetic field} 

\author{Ahmet Kele\c{s}}
\email[]{ahmetkeles99@gmail.com}
\affiliation{Department of Physics and Astronomy, University of Pittsburgh,
  Pittsburgh, Pennsylvania 15260, USA}
\affiliation{School of Physics and Astronomy and Computational Sciences,
  George Mason University, Fairfax, Virginia 15260, USA}

\author{M. \"{O}. Oktel}
\email[]{oktel@fen.bilkent.edu.tr}
\affiliation{Department of Physics, Bilkent University, 06800, Ankara, TURKEY}

\begin{abstract} 
  We consider the Bose-Hubbard model on a two-leg ladder under an artificial
  magnetic field, and investigate the superfluid to Mott insulator transition
  in this setting.  Recently, this system has been experimentally realized
  [M.Atala \emph{et al.}, Nature Physics {\bf 10}, 588--593 (2014)], albeit in
  a parameter regime that is far from the Mott transition boundary. Depending
  on the strength of the magnetic field, the single-particle spectrum has
  either a single ground state or two degenerate ground states. The transition
  between these two phases is reflected in  the many particle properties. We
  first investigate these phases through the Bogoliubov approximation in the
  superfluid regime and calculate the transition boundary for weak
  interactions. For stronger interactions the system is expected to form a
  Mott insulator. We calculate the Mott transition boundary as a function of
  the
  magnetic field and interleg coupling with  mean field theory,  strong
  coupling expansion, and  density matrix renormalization group (DMRG).
  Finally, using the DMRG, we investigate the particle-hole excitation gaps of
  this system at different filling factors  and find peaks at simple fractions
  indicating the possibility of correlated phases.  
\end{abstract} 

\pacs{ 05.30.Jp, 05.70.Fh, 67.25.dj,03.75.Lm, 73.43.-f,05.10.-a } 

\maketitle

\section{Introduction}
\label{sec:intro}

Cold atom experiments can realize the fundamental models of many particle physics
which are not accessible with traditional condensed matter techniques.  One
recent advance has been the demonstration of artificial magnetic fields in
optical lattice systems, as well as in continuums
\cite{spielman,ketterle,blochlattice}.  The optical lattice experiments
control the phase of the hopping between lattice sites to create a
Hamiltonian with an artificial magnetic field.  This effective magnetic field
is orders of magnitude larger than what is attainable in a solid state
experiment. For the typical lattice constants in solids, the magnetic flux
through a unit cell is comparable to flux quantum $h/e$ only for magnetic
fields in excess of thousands of teslas. The first experiments demonstrating
effective magnetic fields in optical lattices have proven that this
extremely high magnetic field regime is accessible with cold atoms
\cite{ketterle,blochlattice}.

Most investigations of magnetic field effects in many particle systems
rely on a separation of length scales, assuming that the magnetic length is
much larger than the lattice scale. However, if these two length scales are
comparable, the magnetic field can no longer be treated semiclassically and
has to be directly taken into account in the microscopic Hamiltonian. The
profound effect of such strong magnetic fields can be observed even for
non-interacting particles. The single-particle spectrum is sensitively
dependent on the external field, forming a self-similar structure known as the
Hofstadter butterfly \cite{hofstadteroriginal}. Recent experiments hold
the promise for investigation of many-particle physics for systems with such
complicated single-particle dispersions. The interplay between interactions and
the complicated single-particle spectrum is expected to result in novel phases
\cite{onur,novelphases}.

The first experiments which implemented an artificial magnetic field for
lattice systems demonstrated the existence of the artificial magnetic field by
measuring the effect of this field on  excited states of the system
\cite{ketterle,blochlattice}.  Thus they did not probe the ground state of
the Hofstadter-Hubbard Hamiltonian. The recent experiment by the Munich group
has, for the first time, demonstrated the effects of an artificial magnetic
field on the ground state of a lattice system. 

The experiment in Ref. \cite{THEEXPERIMENT} realizes a model which is
essentially one dimensional. In general, the orbital coupling of the magnetic
field to a one dimensional system does not create any change, as such a field
can be set to 0 by a gauge transformation. However, by using a two-leg
ladder, the experiment creates a situation in which the magnetic field has
non-trivial effects on the system without generating a complicated single
particle spectrum or a sensitive dependence on the rationality of the applied
field. Thus, experimental realization of this system provides the first
opportunity to study the behavior of lattice bosons in an extremely high
magnetic field regime.  

In this paper, we investigate this model system theoretically, particularly
focusing on the effect of the  artificial magnetic field on the Mott
insulator--to--superfluid transition. We have previously conducted a theoretical study of
the two-leg Bose-Hubbard ladder \cite{keles2009}. In this paper our
unpublished results are summarized and extended to cover the regime
investigated by the experiment.  

We find that the transition between the Meissner and the vortex phases moves
to a higher magnetic field for weak interactions. For strong interactions the
system goes into the Mott insulator state. We find consistent results from the
strong coupling-expansion and density matrix renormalization group (DMRG) for
the Mott insulator boundary. A magnetic field stabilizes the Mott state and
makes it accessible at a lower interaction strength. We also find that there
is a re-entrant Mott transition as a function of the hopping strength at fixed
chemical potential. Finally, we investigate the gap between the ground and the
first excited states through the DMRG and find that there are distinct peaks
at simple filling fractions, providing evidence for the existence of
correlated states in this system.  

The paper is organized as follows: We introduce the Hamiltonian in
Sec.~\ref{sec:model}, and review the properties of the single-particle spectrum
in Sec.~\ref{sec:singlepart}.  In Sec.~\ref{sec:gp}, we investigate
the system with weak interactions using the Gross-Pitaevskii mean-field
approximation and also discuss the excitations of the system above the mean
field solution. The remaining sections focus on the strongly interacting
regime. In Sec.~\ref{sec:meanfield}, we calculate the phase diagram of
the system using a real space Gutzwiller ansatz. This approximation is
particularly poor for one dimensional systems, thus in Sec.~\ref{sec:sce}
we calculate the phase diagram using  strong coupling perturbation theory.
Sec.~\ref{sec:dmrg} contains the discussion of  the Mott transition using
the DMRG. In Sec.~\ref{sec:correlated}, we investigate the gap between the
ground and the first excited state of the system at half-filling in the
infinite interaction limit as well as the gaps in the particle-hole
excitations of the system for various fillings using the DMRG and discuss the
possibility of correlated states.  Finally, we summarize our results and their
consequences for  experiments in Sec.~\ref{sec:concl}.

\section{Model} 
\label{sec:model}

We consider an infinite ladder composed of square plaquettes extending in the
$\hat{x}$ direction, with nearest-neighbor hopping. The tight-binding
Hamiltonian for this two-leg ladder is given by
\begin{eqnarray}
  H=
  &-& \sum_i \left[
    Je^{-i\alpha}a_i^\dagger a_{i+1}+
    Je^{i\alpha}b_i^\dagger b_{i+1} +
    Ka_i^\dagger b_i  +H.C.\right] \nonumber\\ 
  &+&\frac{U}{2}\sum_i 
    n^a_i(n^a_i-1)+
    n^b_i(n^b_i-1)-
    \mu\sum_in^a_i+n^b_i,
    \label{eq:hamiltonian}
\end{eqnarray}
where $a_i$, $b_i$ ($a_i^\dagger$, $b_i^\dagger$) are bosonic annihilation
(creation) operators for the $i$th site in the upper and lower legs,
respectively.  $n_i^a=a_i^\dagger a_i$ and $n_i^b=b_i^\dagger b_i$ are
the corresponding number operators, $J$ ($K$) is the intraleg (interleg) hopping
strength, $U$ is the on-site interaction strength, and $\mu$ is the chemical
potential. We assume a homogeneous system that has ``up-down" symmetry for
zero magnetic field, so that on-site interactions and chemical potentials are
identical for each leg.  The phase $\alpha$ accumulated by hopping from
$\mathbf{r}_i$ to $\mathbf{r}_j$ is 
\begin{equation}
  \alpha=
  \frac{e}{\hbar}
  \int_\mathbf{r_i}^\mathbf{r_j}d\mathbf{r}\cdot\mathbf{A(r)},
  \label{eq:alpha}
\end{equation}
where  $\mathbf{A}$ is the vector potential satisfying
$\nabla\times\mathbf{A}=\mathbf{B}$ and $\mathbf{B}$ is the magnetic field
perpendicular to the two-leg plane. We use  the  Landau gauge
$\mathbf{A}=-By\hat{x}$ for $\mathbf{B}=B\hat{z}$, and choose $y=0$ to be at
the center of two legs so that the upper and lower legs will be at positions
$y=c/2$ and $y=-c/2$, respectively. Thus, the exponent in
Eq.~(\ref{eq:hamiltonian}) can be calculated from Eq.~(\ref{eq:alpha}) as
$\alpha=\pi\phi/\phi_0$, where $\phi$ is the magnetic flux passing through each
plaquette and $\phi_0=h/e$ is the flux quantum. 

The main advantage of considering a two-leg ladder as opposed to a two
dimensional extended system is immediately obvious. For two-dimensional
systems, periodicity under translations can only be obtained when
$\phi/\phi_0$ is taken to be a rational number $p/q$. Only then can the
symmetry broken by the specific gauge choice be restored in a $q$-fold
enlarged unit cell. The two-leg ladder system does not require such a
constraint so that calculations can be carried out for any real number
$\alpha/\pi$ between 0 and 1. As such, the two-leg system presents an
opportunity to observe the non-trivial effects of an external field in a
lattice system without the added theoretical complication. The profound
effect of the magnetic field is evident even at the single-particle level,
which is presented in the next section.

\section{Single Particle Solution} 
\label{sec:singlepart}

\begin{figure}
  \centering 
  \includegraphics[width=0.5\textwidth]{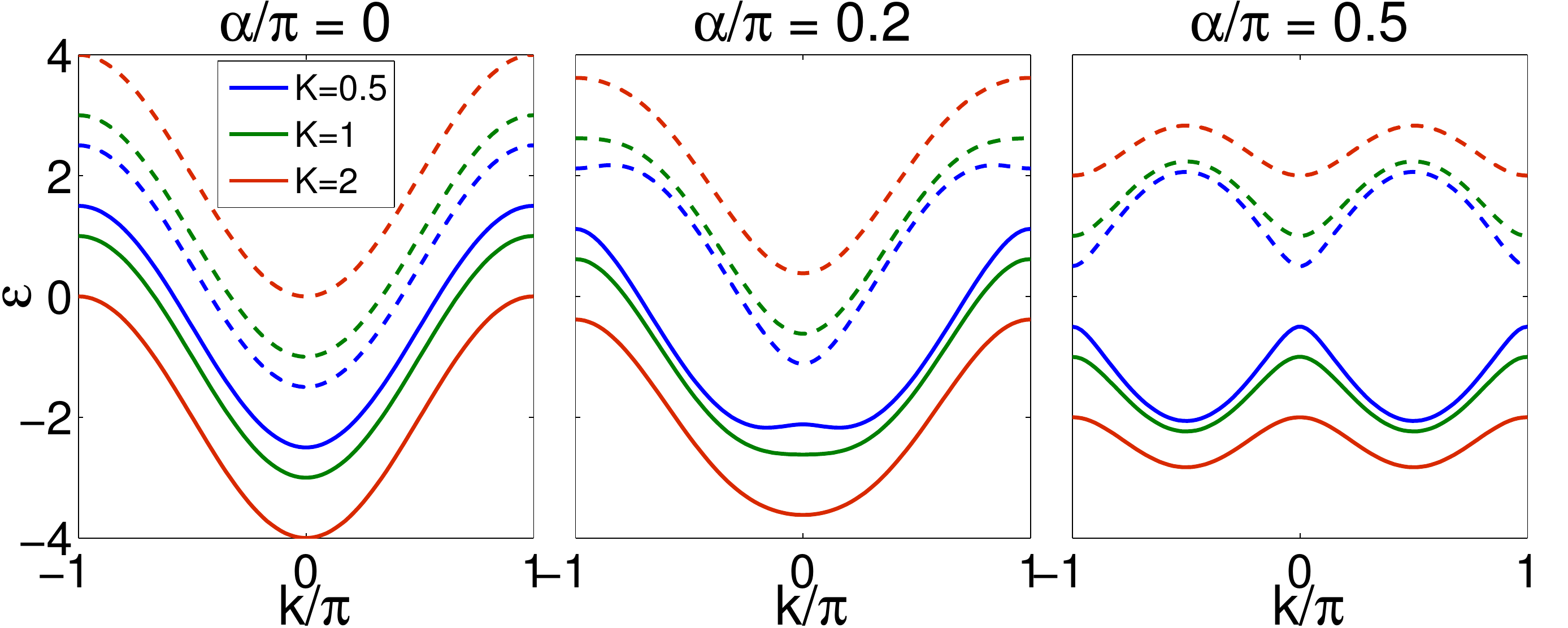}
  \caption{(Color online) Single-particle spectrum of a two-leg ladder with
    varying magnetic flux $\alpha$ and  interleg-to-intraleg hopping ratio
    $K$. Lower bands are shown by solid lines, whereas upper bands are shown
    by dashed lines. The bottom (red) solid line is for $K=2$, the middle
    (green) solid line is for $K=1$, and the top (blue) solid line is for
    $K=0.5$. The gap between lower and higher bands that appears for
    $\alpha/\pi=0.5$ is further shown for the case $K=1$ as a function of the
    magnetic field in Fig.~\ref{fig:hof}.  It is also observed that the gap
    around $k=0$ for $\alpha/\pi=0.5$ closes for very small $K\ll 1$, giving
    rise to a linear dispersion around $k=0$.} 
  \label{fig:dispersion}
\end{figure}

We first give solutions for non-interacting particles; $U=0$. Using the
translational invariance along the $\hat{x}$ direction, the Fourier components of the
field operators can be written as, 
\begin{equation}
  a_j = \frac{1}{\sqrt{L}}\sum_{k} a_k e^{ikj},
  b_j = \frac{1}{\sqrt{L}}\sum_{k} b_k e^{ikj},
  \label{eq:fourier}
\end{equation}
where the Fourier components satisfy the commutation
$[a_k,a_{k'}^\dagger]=\delta_{kk'}$ and $[b_k,b_{k'}^\dagger]=\delta_{kk'}$,
all other commutators being 0. For simplicity, we have taken $c=1$ above so
that all lengths are measured in units of the lattice constant. Using these
transformations in Eq.~(\ref{eq:hamiltonian}), the following Hamiltonian can be
obtained in the momentum space
\begin{equation}
  H_{sp} =
  -\sum_k \left[
  \xi_{ak}a_k^\dagger a_k+
  \xi_{bk}b_k^\dagger b_k +
  Ka_k^\dagger b_k +
  Kb_k^\dagger a_k\right], 
  \label{eq:Hamiltonian2}
\end{equation}
where $\xi_{ak}$ and $\xi_{bk}$ are $2J\cos{(k-\alpha)}$ and
$2J\cos{(k+\alpha)}$, respectively. Diagonalization is achieved by the
Bogoliubov transformation $A_k=\cos{\theta}a_k+\sin{\theta}b_k$,
$B_k=-\sin{\theta}a_k+\cos{\theta}b_k$, where
$\theta=\frac{1}{2}\arctan{(\frac{2 K}{\xi_{ak}-\xi_{bk}})}$. The energy
eigenvalues $\epsilon_{1,2}$ can be found as
\begin{equation}
  \epsilon_{1,2} =
  -2\cos(k)\cos(\alpha)
  \mp\sqrt{\tilde K^2+4\sin^2k\sin^2\alpha },
  \label{eq:dispersion}
\end{equation}
where $\tilde K = K/J$ and we normalize the energy with the interleg hopping
$J$. In Fig.~\ref{fig:dispersion}, we show the dispersion relation in the
first Brillouin zone, for zero and nonzero magnetic fields.  It can be seen
that, as the strength of the field increases, the band minimum in the dispersion
shifts from $k=0$ to two nonzero $k$ values that are degenerate and symmetric
around the origin. The critical field for this bifurcation depends on the
parameter $\tilde K$ as 
\begin{equation}
  \alpha_c = \cos^{-1}
  \left(-\frac{\tilde K}{4} \pm \sqrt{\frac{\tilde K^2}{16}+1}\right).
\end{equation}
Above this critical field, the ground state of the system will no longer be
spatially uniform, but will be a superposition of the plane waves
corresponding to the two minima that can be found from the dispersion as
\begin{equation}
  k_{min}=
  \pm\sin^{-1}\sqrt{\sin^2\alpha-\frac{\tilde K^2}{4\tan^2\alpha}}.
  \label{eq:kmin}
\end{equation} 
In the Munich experiment \cite{THEEXPERIMENT}, these two ground
states were observed for weakly interacting bosons and have been named the
Meissner and vortex phases, respectively.  

\begin{figure}
  \centering
  \includegraphics[scale=0.4]{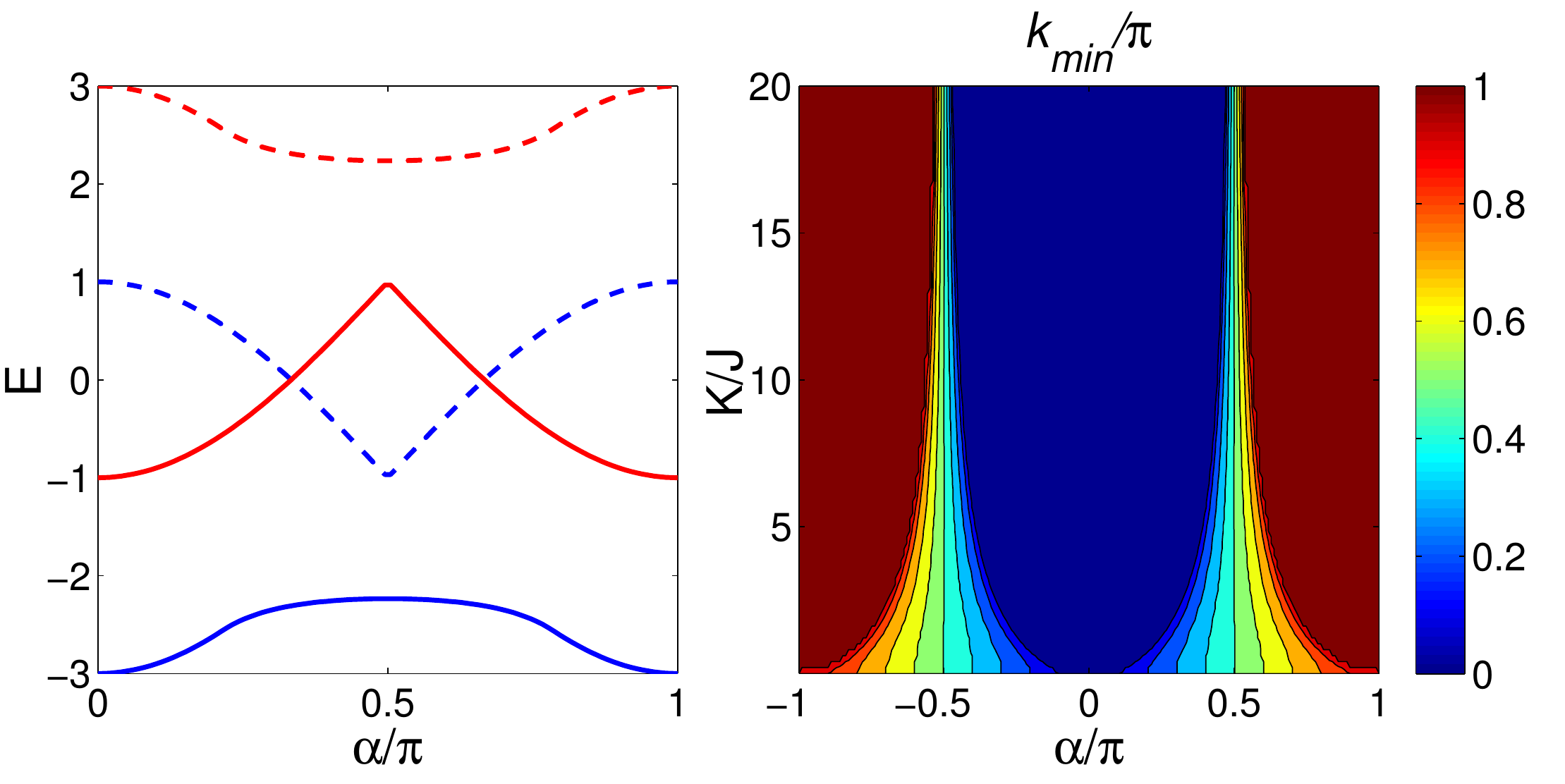}
  \caption{(Color online) Left: Minima and maxima for the two bands as a
    function of the magnetic field. Lower solid and lower dashed (blue) lines
    are the band minimum and maximum of the lower band,
    respectively. Similarly, upper solid and upper dashed (red) lines are the
    minimum and maximum of the the higher band. The band gap is evident between
    $\alpha/\pi=1/3$ and $\alpha/\pi=2/3$, which attains its maximum value at
    $\alpha/\pi=1/2$. Right: Value of the reciprocal lattice
    vector at the minimum energy as a function of the magnetic field and the
    hopping parameter $K/J$.}
  \label{fig:hof}
\end{figure}

As shown in Fig.~\ref{fig:dispersion}, for small values of the
magnetic field, there is no gap between the lower band and the upper band,
whereas for $\alpha/\pi=0.5$ there is a finite band gap between these
two and it decreases as $K/J$ is reduced.  We observe that this gap closes as
$K/J\rightarrow 0$ and a singular  point emerges at $k=0$ in this limit.  To
show more detailed behavior of the band gap, we  plot the minimum and the
maximum of the two bands as a function of the magnetic field for $K=J$ in
Fig.~\ref{fig:hof}. This plot can be regarded as the ``Hofstadter
butterfly'' of the two-leg ladder system. We see that a diamond-shaped gapped
region starts at $\alpha/\pi=1/3$, takes its maximum value $2J$ at
$\alpha/\pi=1/2$, and ends at $\alpha/\pi=2/3$.  In Fig.~\ref{fig:hof}, we
also provide the value of the reciprocal lattice vector $k_{min}$ as given in
Eq.~(\ref{eq:kmin}) at the band minimum as a function of the magnetic field and
the parameter $K/J$, which is in agreement with \cite{THEEXPERIMENT}.

\section{Gross-Pitaevskii Approximation}
\label{sec:gp}

Our picture of the transition between the Meissner and the vortex phases in the
previous section depended only on the non-interacting single-particle spectrum
of the two-leg ladder.  Before we discuss the effects of strong
interactions and the resulting insulating phase we concentrate on the weakly
interacting limit and calculate how the Meissner--to--vortex transition is
affected by the presence of interactions.  

For  small values of the interaction strength and the magnetic field, the
system will essentially be in the superfluid state, mostly dominated by the
hopping term in the Hamiltonian.  Thus, assuming that the condensate
fluctuations are negligible, we make the following approximation: 
\begin{equation}
  a_i \rightarrow \langle a_i\rangle=\psi_i,\quad
  b_i \rightarrow \langle b_i\rangle=\phi_i.
  \label{eq:order}
\end{equation}
Both the amplitude and the phase of those classical fields are time and
position dependent. Clearly, approximation with a uniform condensate will fail
above the critical field.

Making substitution (\ref{eq:order})  in Eq.~(\ref{eq:hamiltonian}), the
following energy functional is obtained (here we take $J=1$ so that $U$,
$\mu$,
and $K$ are in units of $J$):
\begin{eqnarray}
  E=
  &-&\sum_j\left[
    e^{-i\alpha}\psi_j^*\psi_{j+1}+
    e^{i\alpha}\phi_j^*\phi_{j+1}+
    K\psi_j^*\phi_j +
    c.c.\right]\nonumber\\
  &+&\frac{U}{2}\sum_j\left[
    \psi_j^*\psi_j(\psi_j^*\psi_j-1)+
    \phi_j^*\phi_j(\phi_j^*\phi_j-1)\right]\nonumber\\
  &-& \mu\sum_j |\psi_i|^2+|\phi_i|^2.
  \label{eq:energyfunctional}
\end{eqnarray}
Variation of the energy functional around the minimal solutions
$i\partial\psi_i/\partial t=\delta E/\delta{\psi_i}^*$ and
$i\partial\phi_i/\partial t=\delta E/\delta{\phi_i^*}$ gives the following
coupled Gross-Pitaevskii equations: 
\begin{align}
  i\frac{\partial\psi_j}{\partial t} =
    &-\left[ e^{-i\alpha}\psi_{j+1}+K\phi_j+
      e^{i\alpha}\psi_{j-1}\right]
    \nonumber \\
    &+ U|\psi_j|^2\psi_j-(\frac{U}{2}+\mu)\psi_j \\
  i\frac{\partial\phi_j}{\partial t} =
    &-\left[ e^{i\alpha}\phi_{j+1}+K\psi_j+
    e^{-i\alpha}\phi_{j-1}\right]
    \nonumber \\
    &+ U|\phi_j|^2\phi_j-(\frac{U}{2}+\mu)\phi_j.
  \label{eq:grosspitaevskii}
\end{align}
\begin{figure}
  \centering 
  \includegraphics[scale=0.4]{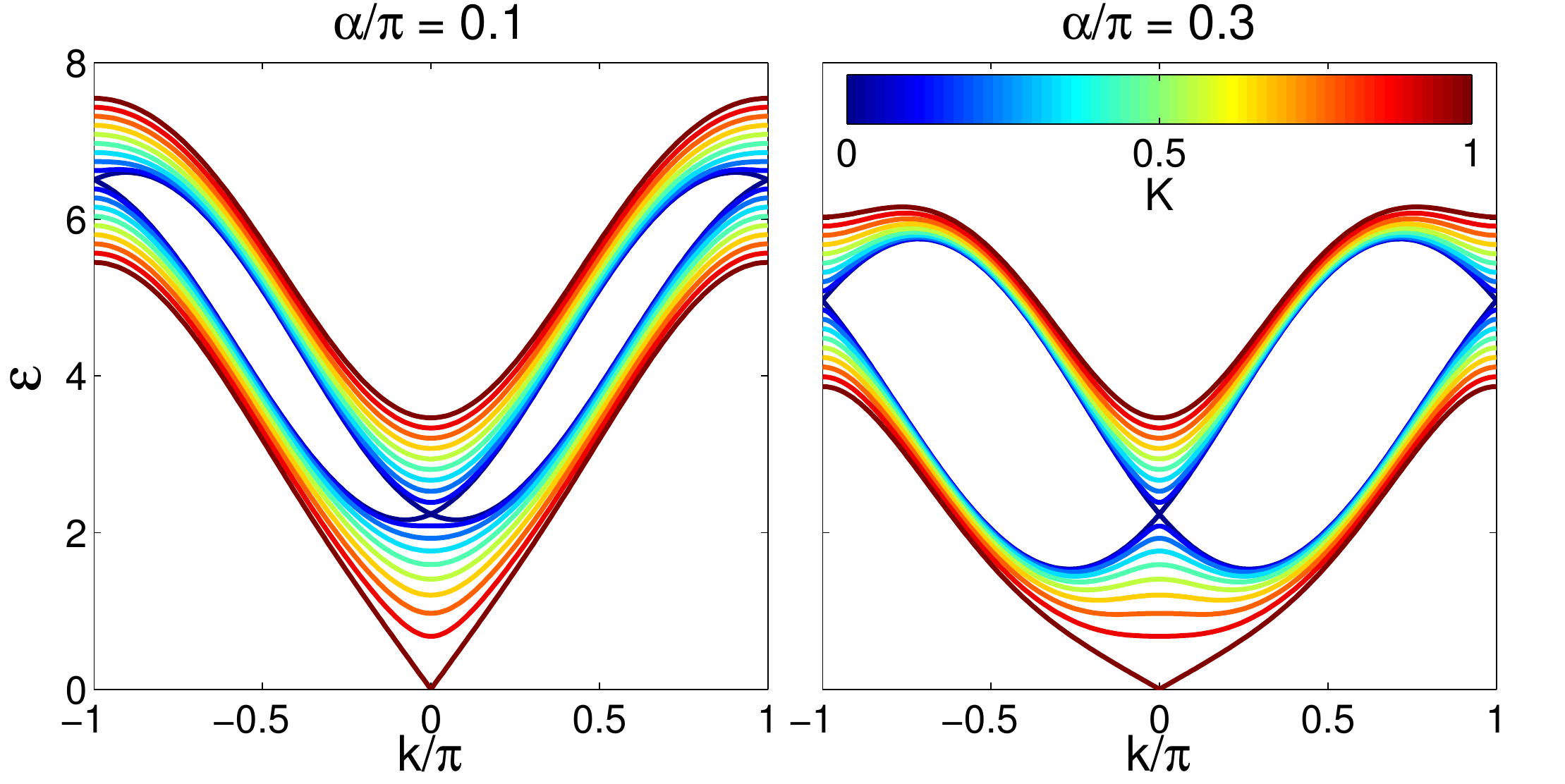}
  \caption{(Color online) Band diagrams for a two-leg ladder with on-site
    interactions calculated within the Gross-Pitaevskii approximation for
    $U=2$. Left: $\alpha/\pi=0.1$. Right: $\alpha/\pi=0.3$. } 
  \label{fig:gpdispersion}
\end{figure}

Zeroth-order terms $\psi_j=\phi_j=\sqrt{n}$  give the chemical potential as
$\mu=-(2\cos{\alpha}+\tilde K)+0.5U(2n-1)$. For a higher order approximation,
the fluctuations in the condensate are taken into account as \cite{landau}:
\begin{align}
  \psi_j&=
    \sqrt{n}+A e^{i(kx_j-\omega t)}
    +B^*e^{-i(kx_j-\omega t)}, \nonumber\\ 
  \phi_j&=
    \sqrt{n}+C e^{i(kx_j-\omega t)}
    +D^*e^{-i(kx_j-\omega t)},
  \label{eq:wavefunctions}
\end{align}
where $A$, $B$, $C$, and $D$ are small complex parameters, ${x_j}$ is the position
of the lattice site and $k$ is the reciprocal lattice vector.  Inserting these
wave functions into Eq.~(\ref{eq:grosspitaevskii}), the equation of motion can be
reduced to an algebraic equation of the form $H_{gp}\vec\Psi=\omega \vec\Psi$
where $\vec\Psi=(A, B, C, D)$ and $H_{gp}$ has the form
\begin{align} 
  H_{gp}=\left[ 
  \begin{array}{cccc}
    -\xi'_{ak}    & Un            & -K            & 0   \\
    -Un           & \xi'_{bk}     & 0             & K   \\
    -K            & 0             & -\xi'_{bk}    & Un  \\
    0             & K             & -Un           & \xi'_{ak}
    \end{array} 
  \right],
\end{align} 
where $\xi'_{ak}=2cos(k-\alpha)-2cos(\alpha)-Un-K$  and
$\xi'_{bk}=2cos(k+\alpha)-2cos(\alpha)-Un-K$. The resulting change in the
spectrum can be obtained by calculating the eigenvalues of $H_{gp}$ which is
shown in Fig.~\ref{fig:gpdispersion}.

\begin{figure}
  \centering 
  \includegraphics[scale=0.45]{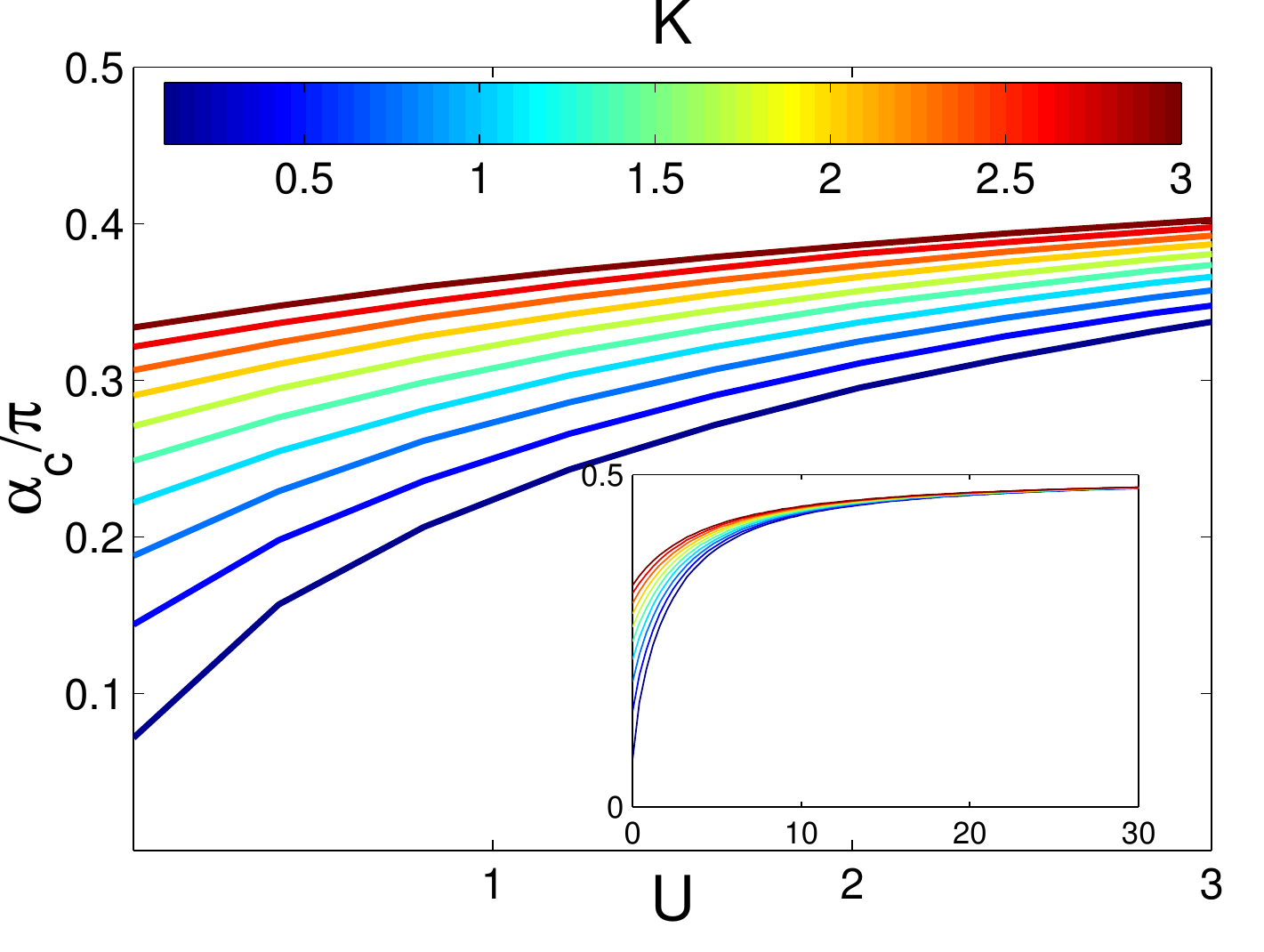}
  \caption{(Color online) Critical magnetic field plotted as a function  of
    the interaction strength $U$ and the intraleg hopping $K$. Inset: The same
    plot zoomed out to span large interactions. Note that the Gross-Pitaevskii
    approximation is not expected to be reliable for strong interactions. }
  \label{fig:alphac}
\end{figure}

Competition between the magnetic field and the interactions can be seen by
considering the band minima around $k=0$ in Fig.~\ref{fig:gpdispersion}.
Here the interactions sharpen the band and provide a cusp-like shape, whereas
the increase in the magnetic field makes it smoother. 

The expansion of the wave function in Eq.~(\ref{eq:wavefunctions}) fails above
the critical magnetic field, as the ground state is no longer spatially
uniform, and the eigenvalues of $H_{gp}$ starts having imaginary parts.  We
have used this property to determine the change of the critical field with the
interaction strength. In Fig.~\ref{fig:alphac}, the critical magnetic field
as function of  the strength of the interaction is shown. It can be seen that,
$U-\alpha_c$ relation is almost linear for small interaction strength but
saturates for strong interactions. It must be noted that for strong
interactions Gross Pitaevskii approximation is not reliable.

\section{Variational Mean Field Approach}
\label{sec:meanfield}

\begin{figure*}
  \centering 
  \includegraphics[scale=0.5]{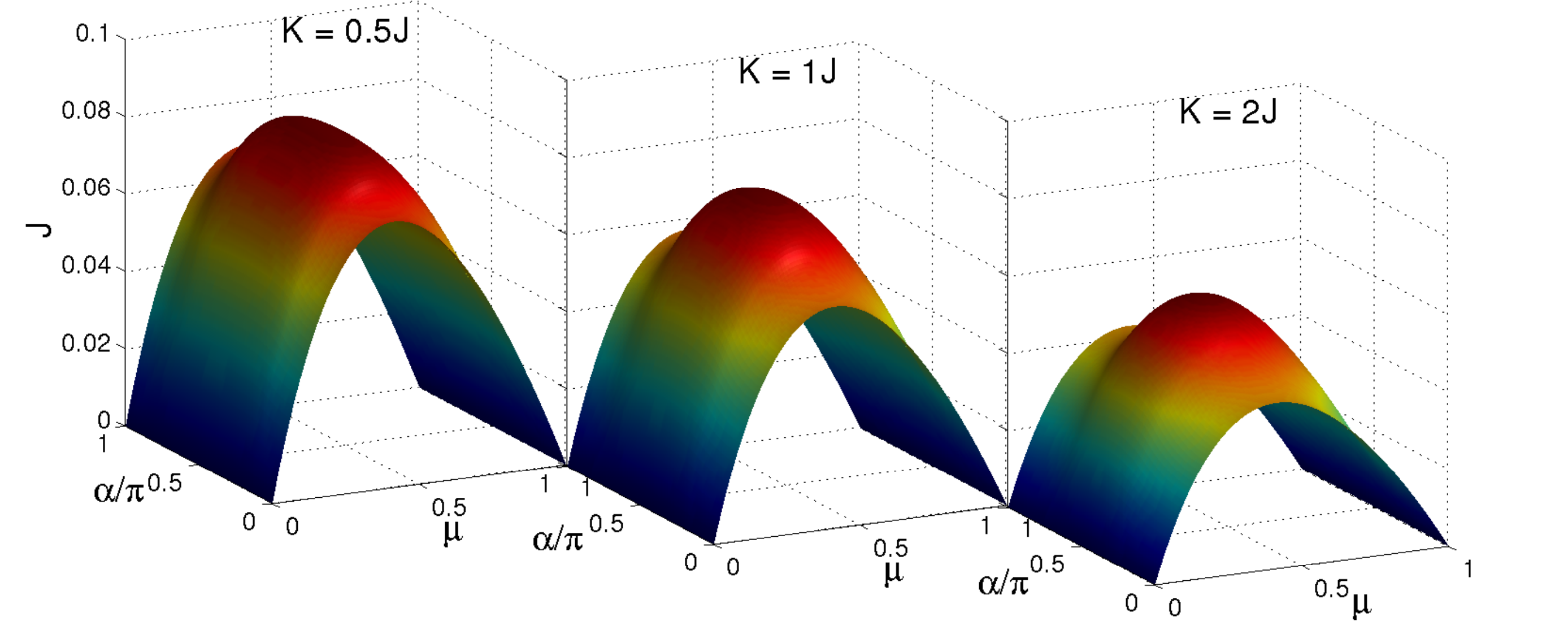}
  \caption{(Color online) Mott insulating phase boundary calculated within
    the variational mean-field approach as a function of the magnetic-field
    strength and the chemical potential for the parameters $K=0.5J$, $K=J$,
    $K=2J$. The region below (above) the plotted surface is the
    insulating (superfluid) state.} 
  \label{fig:mottlobe}
\end{figure*}

In this section, we consider the transition from the superfluid state to the
Mott insulating state as a function of  $J$, $K$, $\mu$ and $\alpha$. Here, it
is convenient to scale the Hamiltonian in Eq.~(\ref{eq:hamiltonian}) with
$U=1$. In the perfect Mott insulator phase, each site has a localized wave
function with exactly $n_0$ particles such that the wave function in each site
is $|n_0\rangle_i$ in the Fock basis.  Allowing small variations around this
equilibrium, we write the following Gutzwiller ansatz for local sites;
\begin{align} 
  |G\rangle_{ak}
    &=\Delta_{ak}|n_0-1\rangle_{ak}
    +|n_0\rangle_{ak}
    +\Delta'_{ak}|n_0+1\rangle_{ak},
    \nonumber\\
  |G\rangle_{bk}
    &=\Delta_{bk}|n_0-1\rangle_{bk}
    +|n_0\rangle_{bk}
    +\Delta'_{bk}|n_0+1\rangle_{bk},
  \label{eq:gutzwiller}
\end{align}
where $\Delta$ and $\Delta'$ are small complex variational parameters.
Wavefunction for a rung is $|G\rangle_{rk}=|G\rangle_{ak}|G\rangle_{bk}$ so
that the total wavefunction of the system can be written as
$|\Psi\rangle=\prod_{k}|G\rangle_{ak}|G\rangle_{bk}$. The variational energy of
the system is calculated from $\varepsilon=\langle\Psi|H|\Psi\rangle/\langle
\Psi|\Psi\rangle$ up to second order in $\Delta$ and $\Delta'$ as follows:
\begin{widetext}
  \begin{eqnarray}
     \varepsilon =\sum_{i=1}^{N}
       \{
     &-&Je^{-i\alpha}
     \left[
       n_0\Delta_{a,i}\Delta^*_{a,i+1}
       +(n_0+1)\Delta'^*_{a,i}\Delta'_{a,i+1} 
       +\sqrt{n_0(n_0+1)}\Delta_{a,i}\Delta'_{a,i+1} 
       +\sqrt{n_0(n_0+1)}\Delta'^*_{a,i}\Delta^*_{a,i+1} 
     \right]
     -Je^{i\alpha}
     \left[
       a\rightarrow b
     \right] \nonumber\\
     &-&K
     \left[
       n_0\Delta_{a,i}\Delta^*_{b,i}
       +(n_0+1)\Delta'^*_{a,i}\Delta'_{b,i}
       +\sqrt{n_0(n_0+1)} 
        (\Delta_{a,i}\Delta'_{b,i}+\Delta'^*_{a,i}\Delta^*_{b,i}) 
     \right] \nonumber\\
     &+&
     \left[
       (1-n_0+\mu)(|\Delta_{a,i}|^2+|\Delta_{b,i}|^2)
       +n_0(n_0-1-2\mu)
       +(n_0-\mu)(|\Delta'_{a,i}|^2+|\Delta'_{b,i}|^2)
     \right]
     \}.
     \label{eq:gutzwillerenergy}
   \end{eqnarray}
We minimize the energy with respect to  $\Delta_{ai}$, $\Delta_{bi}$,
$\Delta'_{ai}$, and $\Delta'_{bi}$. The Jacobian matrix of the second
derivatives is calculated as 
\begin{eqnarray}
 \mathbf{J}=-
 \left(
   \begin{array}{cc}
     n_0\mathbf{F}                & \sqrt{n_0(n_0+1)}\mathbf{F} \\
     \sqrt{n_0(n_0+1)}\mathbf{F}  & (n_0+1)\mathbf{F} 
   \end{array}
 \right) 
 +
 \left(
   \begin{array}{cc}
     (1-n_0+\mu)\mathbf{I} &         0        \\
             0           & (n_0-\mu)\mathbf{I} 
   \end{array}
 \right),
\label{eq:jacobian}
\end{eqnarray}
where $\mathbf{I}$ is a $2N\times 2N$ identity matrix and $\mathbf{F}$ is
written as
\begin{equation}
  \mathbf{F}=\left[
    \begin{array}{cccccc}
      A         & B         & \ldots    & B^\dagger\\
      B^\dagger & A         & \ddots    & 0\\
      \vdots    & \ddots    & \ddots    & B\\
      B         & 0         & B^\dagger & A
    \end{array}
  \right].
  \label{eq:tmatrix}
\end{equation}
Here sub-blocks are defined in terms of Pauli matrices in the upper leg--lower
leg basis as
\begin{align}
  A = K\sigma_x, \quad 
  B = Je^{i\alpha\sigma_z}.
\end{align}
To find the eigenvalues, we use the same method presented in
Ref.~\onlinecite{onur}: Let $\lambda_\mathbf{F}$ and $\vec{u}$ be the eigenvalues
and the eigenvectors of $\mathbf{F}$, respectively: then one can apply an ansatz of
the form 
$\vec{v}=(
   a\vec{u},
   b\vec{u} 
)$
and solve the eigenvalue equation $\mathbf{J}\vec{v}=\lambda\vec{v}$, which is
found to be
\begin{equation}
 \lambda_{1,2}
  =1-\lambda_\mathbf{F}(2n_0+1)
  \pm\sqrt{(1-\lambda_\mathbf{F}(2n_0+1))^2+4\lambda_\mathbf{F}
  (\mu+1)-4(n_0-\mu)(1-n_0+\mu)}.
  \label{eq:lambda12}
\end{equation}
\end{widetext}
Equating the minimum eigenvalue of the Jacobian matrix in
Eq.~(\ref{eq:lambda12}) to 0 yields the phase boundary of the Mott
insulating region. Solving the corresponding equation for $K$ and $J$, the
following simple relation can be found for the boundary of the Mott phase
\begin{equation}
  \lambda_\mathbf{F}(K_c,J_c)=\frac{(n_0-\mu)(1-n_0+\mu)}{(\mu+1)}.
  \label{eq:meanfieldmott}
\end{equation}
Here $\lambda_\mathbf{F}$ is the minimum value of $\epsilon_1$ in Eq.
\ref{eq:dispersion} so that we obtain the Mott phase boundary for each value
of the magnetic field $\alpha$.  In Fig.~\ref{fig:mottlobe},
Eq.~(\ref{eq:meanfieldmott}) is plotted for $n_0=1$ which shows the shape of
the Mott insulation region.

Note that this result is exact within  the mean-field theory.  However, the
mean-field theory in a quasi-one-dimensional system is not expected to be
accurate.  The decoupling of the hopping term in Eq.~(\ref{eq:hamiltonian}) by
introducing a mean-field is questionable in a low-dimensional system, where
the
effect of fluctuations is necessarily important.  The mean-field calculation
can only describe the system at a qualitative level.  It provides a general
idea about the topology of the Mott region, and an estimate of the phase
boundary for small values of the hopping strength where site-site correlations
are diminished.  For a better determination of the Mott insulating region, we
turn to more accurate methods in the following sections.

\section{Strong-Coupling Expansion}
\label{sec:sce}

A better description of the transition is obtained by treating the hopping
term as a perturbation in the perfect Mott state. While this is, in spirit,
close to the mean field approach given in the previous section, correlations
between the sites are built in as higher orders in perturbation theory are
developed. The resulting ``strong coupling expansion'' has been successfully
applied to the Bose-Hubbard model in low dimensions and has been shown to be in
perfect agreement with accurate numerical methods \cite{kuhner,kuhner2}.

In the strong coupling expansion, the hopping amplitude is considered as a small
parameter. The Mott insulator state is characterized by a finite gap for
particle-hole excitations, whereas this gap vanishes for the superfluid
phase\cite{fisher}. We calculate the energy of a system with exactly $n_0$
particles per site (Mott state $E_M$) and the energy of a system with one
additional defect (particle $E_P$ or hole $E_H$) perturbatively. The energy
difference between the defect states and the perfect Mott state vanishes at
the phase boundary.  This method has been used for systems with different
dimensions\cite{freericks1,freericks2} and for a two dimensional system under
a magnetic field\cite{monien}.

For calculations under perturbation theory, it is convenient  to write the
Hamiltonian in the generalized form,
\begin{equation}
 H=-\sum_{ij}\mathbf{F}_{ij}\tilde a_i^\dagger \tilde a_j +
\frac{1}{2}\sum_i\tilde n_i(\tilde n_i-1)-\mu\sum_i\tilde n_i
\end{equation}
where $\mathbf{F}$ is given in Eq.~(\ref{eq:tmatrix}) for our model and the
correspondence between the operators $\tilde a_i$ and $\tilde n_i$ and the
operators in Eq.~(\ref{eq:hamiltonian}) is obvious.

We  perform-strong coupling perturbation up to second order in our
calculations. The energies of the Mott state $E_M$, the additional particle
state $E_P$, and the additional hole state $E_H$ are found to be
\begin{align}
  E_M &= 
    E_M^0
    -Nn_0(n_0+1)(2J^2+K^2), \\
  E_P &=
    E_P^0
    -(n_0+1)\lambda_\mathbf{F}
    -Nn_0(n_0+1)(2J^2+K^2)
    \nonumber\\
    &-n_0(n_0+1)\lambda_\mathbf{F}^2
    +\frac{1}{2}n_0(5n_0+4)(2J^2+K^2), \\ 
  E_H &=
    E_H^0
    -n_0\lambda_\mathbf{F}
    -Nn_0(n_0+1)(2J^2+K^2)
    \nonumber\\
    &-n_0(n_0+1)\lambda_\mathbf{F}^2
    +\frac{1}{2}(n_0+1)(5n_0+1)(2J^2+K^2),
    \label{eq:scpenergies} 
\end{align}
where $\lambda_\mathbf{F}$ is the lowest eigenvalue of hopping matrix
$\mathbf{F}$ and $N$ is the number of lattice sites in one leg. Zeroth-order
energies are $E_M^0=2N(n_0(n_0-1)/2-\mu n_0)$, $E_P^0=E_M^0+n_0-\mu$, and
$E_H^0=E_M^0-(n_0-1)+\mu$. Solving the equations $E_P-E_M=0$ and $E_M-E_H=0$
for the chemical potential $\mu$ separately, the phase boundary of the
particle sector and hole sector is obtained as,
\begin{align}
 \mu_P&=
   n_0
  +(n_0+1)\lambda_\mathbf{F}
  -n_0(n_0+1)\lambda_\mathbf{F}^2\nonumber\\
   &+\frac{1}{2}n_0(5n_0+4)(2J^2+K^2), \\
 \mu_H&=
   (n_0-1)
   -n_0\lambda_\mathbf{F}
   -n_0(n_0+1)\lambda_\mathbf{F}^2\nonumber\\
   &-\frac{1}{2}(n_0+1)(5n_0+1)(2J^2+K^2). 
\end{align}
Here the magnetic-field dependence comes indirectly from the eigenvalue
$\lambda_\mathbf{F}$, but higher order terms in the perturbation will depend
on the magnetic field explicitly.  An interesting observation is that our
results to this order are similar to the results in Ref.\onlinecite{monien}
for a number of nearest neighbors  equal to $3$. However, this is not guaranteed
for higher order expansions since the flux attained through hopping is
different due to the difference in the topology of this constrained problem.
The eigenvalue spectrum of the $\mathbf{F}$ matrix is  shown in
Fig.~\ref{fig:hof}. 
\begin{figure}
  \begin{center}
    \includegraphics[scale=0.32]{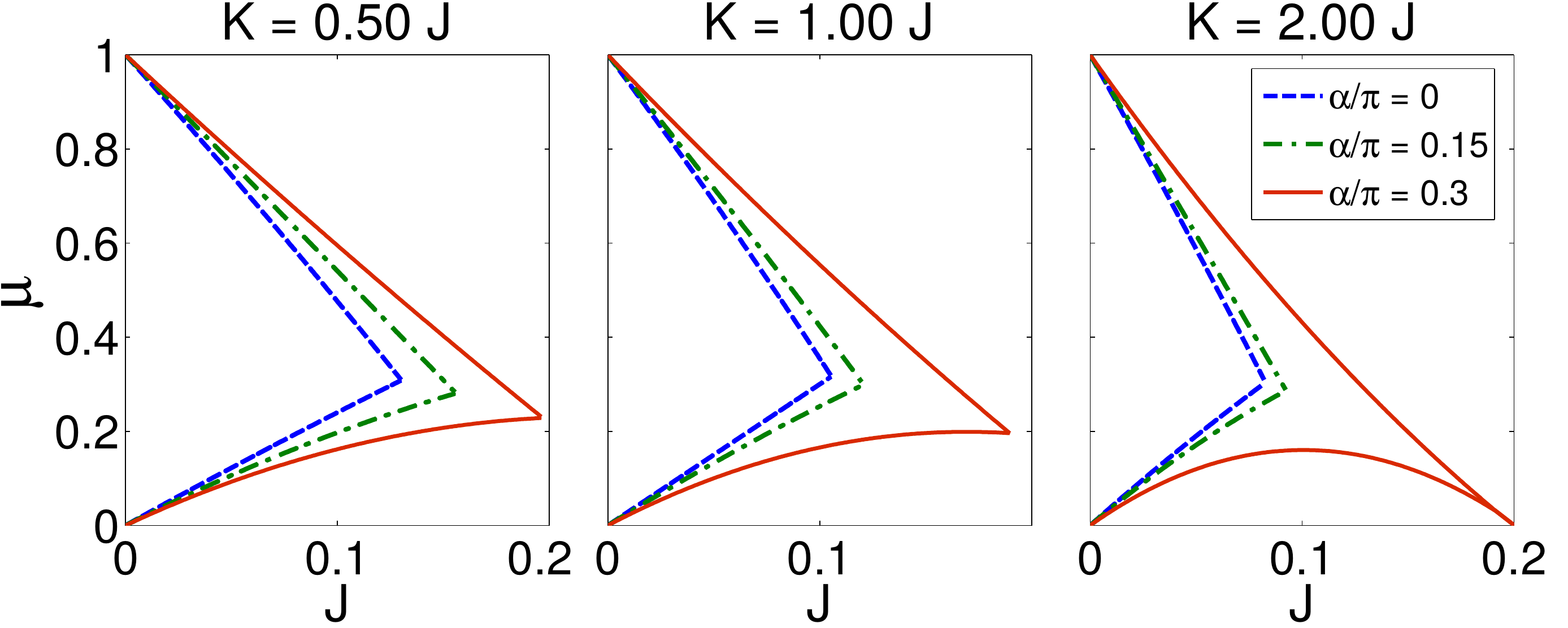}
  \end{center}
  \caption{(Color online) Phase diagram of a two-leg ladder from the
    strong-coupling expansion up to second order for different magnetic
    fields and interleg-to-intraleg hopping ratios.} 
  \label{fig:scp}
\end{figure}

In Fig.~\ref{fig:scp} we show the results of this calculation. An increase
in the magnetic field enlarges the Mott insulating region of the phase
diagram. This is expected as the magnetic field localizes the single-particle
trajectories even for the noninteracting problem thus a transition to an
insulator state is easier. The Mott lobe grows in size until $\alpha=0.5$ and
then reduces to satisfy periodicity at $\alpha=1$.  The shape of the lobe is
not concave as predicted by the mean field, but convex with a cusp at the tip.
This shape is generic in one dimension, as obtained by strong-coupling, Monte
Carlo, and the DMRG results in one dimension. Comparing
Figs.~\ref{fig:mottlobe} and \ref{fig:scp}, it can be observed  that the mean-field results underestimate the Mott boundary by a considerable amount.

A new feature of the phase diagram emerges after $\alpha=0.3$. The Mott phase
has a re-entrance as a function of the hopping strength at a fixed chemical
potential. (Beyond $\alpha=0.3$ for $K=2$, curves of the particle and the hole
sector intersect at such a large value of the hopping amplitude that the
second-order perturbation theory fails to capture this region).  The behavior
of the phase diagram here can be explained as follows.  At fixed chemical
potential, if we start from the pure Mott state  one expects the excitations
above this state to be mobile, as they move on a constant background of filled
sites.  If the kinetic energy gained by this mobility compensates the
interaction energy with the background, these excitations become energetically
favorable and cause the Mott insulator--to--superfluid transition.  The magnetic
field modifies this usual transition mainly by limiting the mobility of the
excitation; this is why, in higher dimensional Bose-Hubbard models, the
magnetic field increases the size of the Mott lobes in the phase diagram.   In
the two-leg ladder this confining effect takes an interesting form: starting
from the pure Mott state, introduction of a small hopping causes the system to
become superfluid, however, further increase in the hopping strength localizes
the excitations and causes the system to go back to the Mott insulating state.
Thus, for small $J$ the excitation energies are not affected by the magnetic
field, but as  hopping is increased this term becomes dominant and causes a
phase transition back to the insulator phase.  As J is increased further the
system is once again dominated by kinetic energy and reaches the superfluid
state.

The re-entrant phase behavior found in one dimensional systems appears in the
two-leg ladder with an increase in the magnetic field.  This re-entrant behavior
was not observed in the results of strong coupling perturbation in one, two, or
three dimensions, or in a two dimensional lattice under a magnetic field (in
\cite{freericks1,freericks2,monien} perturbation was carried out up to 
third order). The existence of this re-entrant phase is also supported by our DMRG
results, which is the subject of the next section.

\section{DMRG Calculations}
\label{sec:dmrg}

DMRG has been proven to provide
numerically exact solutions of one dimensional lattice
systems \cite{white1,white2}.  This method has been extensively applied to the
Bose-Hubbard model \cite{kuhner,kuhner2,pai3} and shown to be one of the most
reliable approaches for quasi-one dimensional systems.  Thus, in this section,
we use the DMRG to calculate the Mott transition boundary for a two-leg Bose-Hubbard ladder under a magnetic field.  

We use a method similar to that in \cite{pai4}, namely, rung by rung
enlargement, but employ single rung enlargement \cite{whitesinglecenter} in
the construction of the superblock Hamiltonian.  We use the finite-system DMRG
algorithm for a ladder of $60$ rungs and for each site we set the maximum
occupancy $n_{max}=4$.  Particle number conservation is used to diagonalize
only the $N_{particle}=N_{sites}$ sector of the superblock Hamiltonian or
$N_{particle}=N_{sites}\pm 1$ as additional target states.  Further details
about the projection to the space with different fillings are given in the
next section.

\begin{figure}
  \begin{center}
    \includegraphics[scale=0.5]{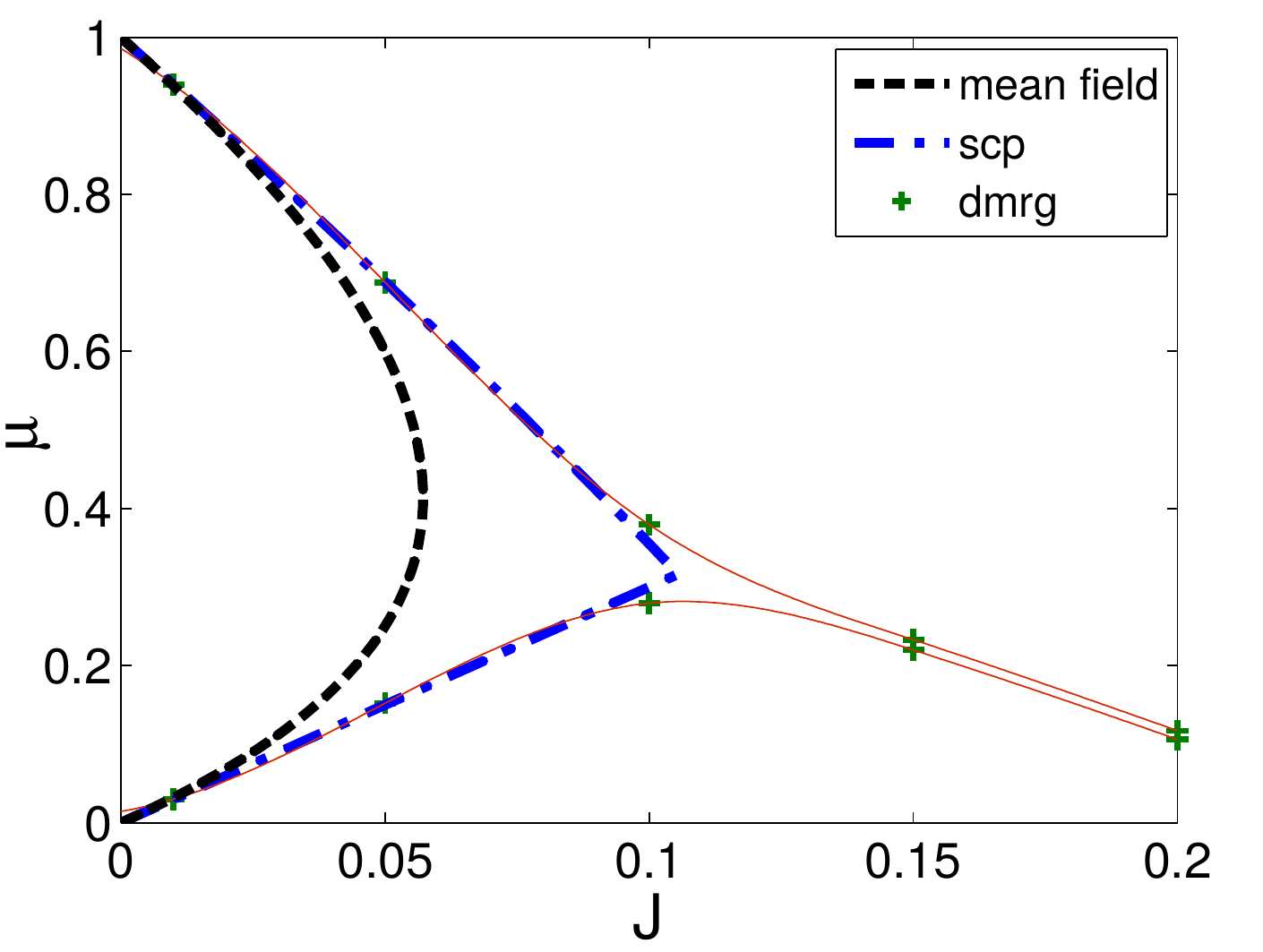}
    \includegraphics[scale=0.5]{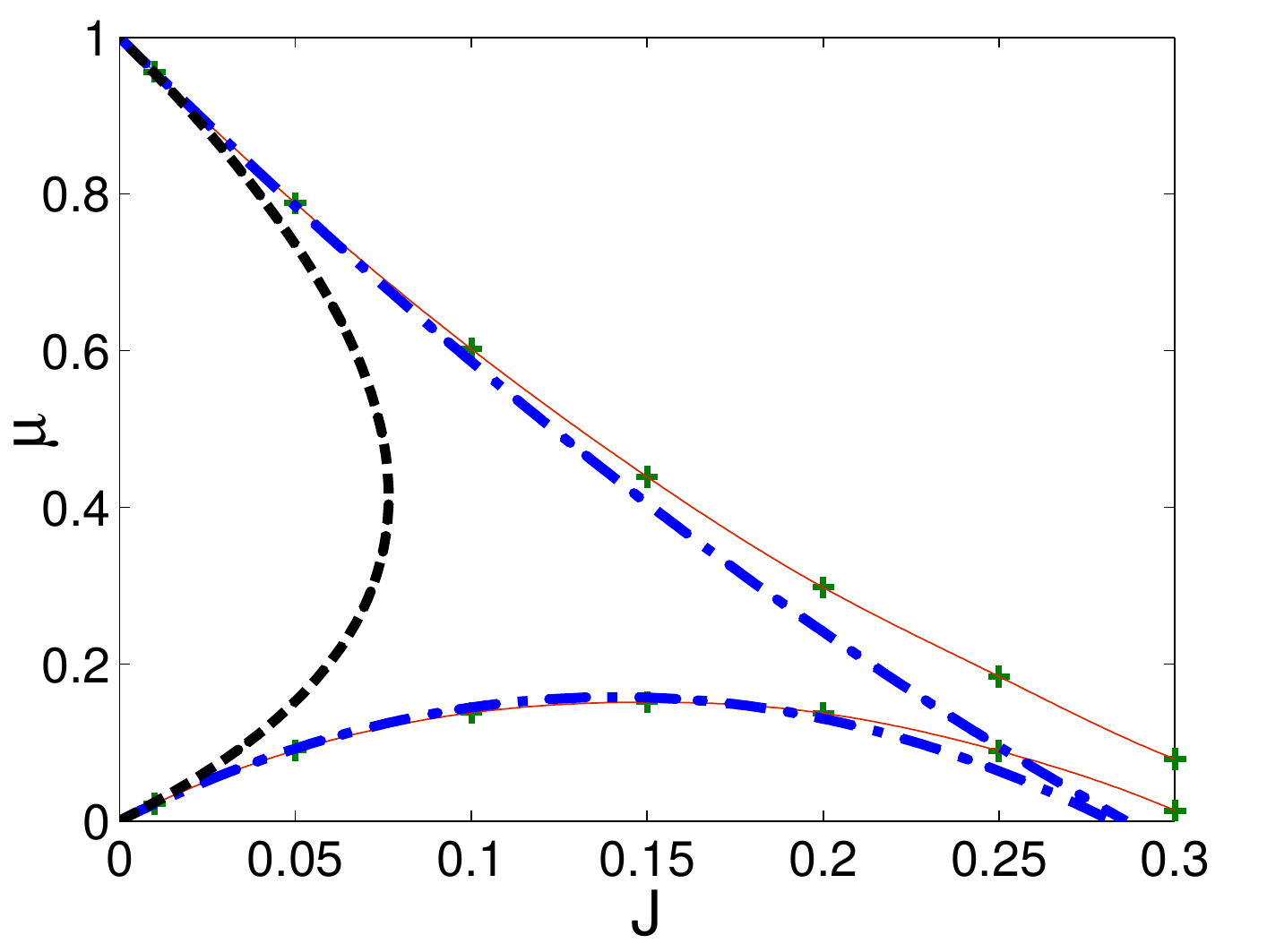} 
  \end{center} 
  \caption{(Color online) Phase diagram of a two-leg Bose-Hubbard ladder
    from the mean-field, strong-coupling expansion and DMRG at $\alpha=0$
    (top) and$\alpha/\pi=0.45$ (bottom). The thin solid (red) line is
    the spline interpolation to DMRG data points. } 
  \label{dmrg}
\end{figure}

Calculation of the  Mott phase boundary via DMRG is very similar to the strong
coupling perturbation method. One needs the energies of the Mott phase together
with the additional particle and hole states to find the phase boundary. The
energies of particle and hole states are calculated as additional target
states in the DMRG implementation \cite{kuhner}.  In Fig.~\ref{dmrg}, one can see
the good agreement between the strong coupling result and the DMRG.  For larger
values of hopping, the strong-coupling deviates from the DMRG, as expected for a
perturbative method. Another point is that the existence of re-entrance is
also validated by DMRG results. We show a similar phase for $\alpha=0.45$ in
Fig.~\ref{dmrg}. It is seen that strong coupling calculations give
relatively poor results above $J\approx 0.2$. 

Finally, we note that the tip of the Mott insulator region requires a special
treatment with DMRG.  The two branches coming from the particle and hole
sector intersect only in the thermodynamic limit, whereas our system is
composed of only $60$ rungs. There are several approaches (like consideration
of the correlation length and extrapolation to the Luttinger liquid
correlation function in \cite{kuhner}) to remedying this situation. As the
critical behavior of the tip is not our main concern in this paper, we do not
perform a similar analysis.

\section{Evidence of Strongly Correlated Phases}
\label{sec:correlated}

In the previous sections, we have performed various calculations that can only
work close to  the Mott insulator phase. Theoretical approaches are limited
for the two-dimensional Bose-Hubbard model under a magnetic field,
particularly for strong fields.  This is due to the complicated single
particle spectrum as well as the interplay between  strong correlations and
the high number of degeneracies.  Both strong-coupling and the mean-field
approaches work in the region where such correlations are weak. On the other
hand, this is exactly the region where novel phases are expected.  For this
reason, characterization of the two-dimensional Bose-Hubbard model exposed to
a strong magnetic field is attracting close attention.  There have been
several proposals that try to connect these strongly correlated states with
the formation of a vortex lattice or with the incompressible quantum liquids
found in quantum Hall effect \cite{novelphases}. The absence of an
encompassing theoretical model makes it difficult to identify the physics of
this regime. 

Both the strong-coupling expansion and the  mean-field theory as discussed in
the previous sections use the Mott insulator state as their starting point. As
a result, their range of validity is limited to densities close to integer
filling. In the other limit, the Gross-Pitaevskii approximation assumes a
uniform gas spread over the lattice to reveal the dynamics of the system.
Compared to these theoretical approaches, the DMRG has a very wide range of
applicability regardless of the particle number, strength of the field, or
interaction.  One can calculate the ground state of the system for a finite
lattice with any number of particles for all values of the magnetic field and
the interaction strength. In this section, we use the DMRG method to study the
two-leg Bose-Hubbard model under a magnetic field outside the Mott insulator
region, and look for evidence of strongly correlated behavior.

\begin{figure*}[t] 
  \begin{center} 
    \includegraphics[scale=0.6]{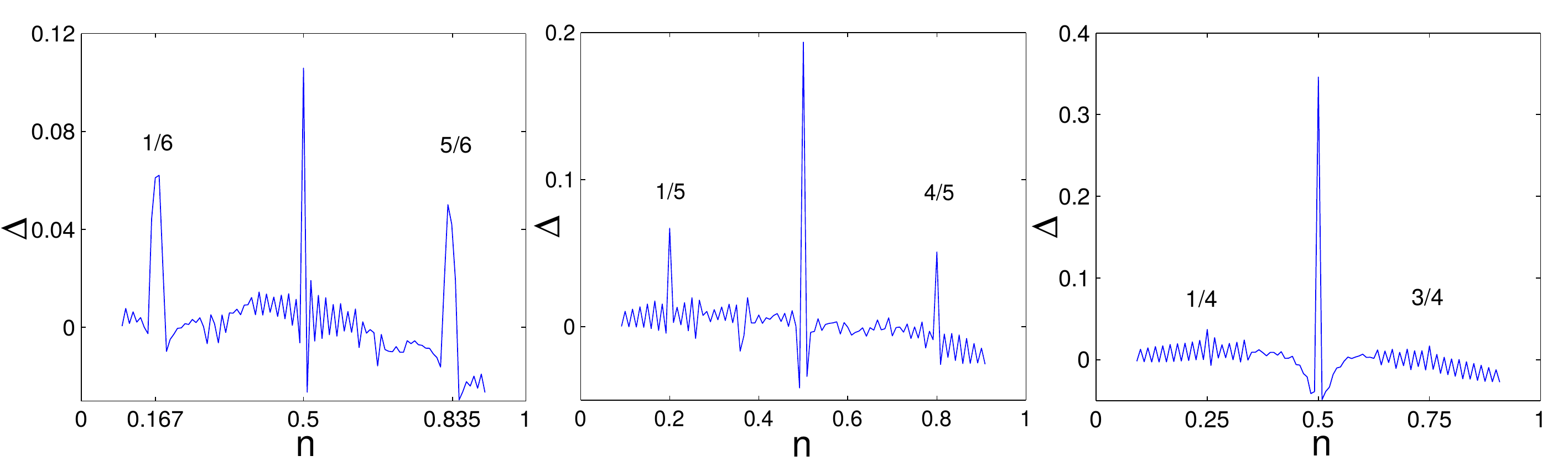}
  \end{center} 
  \caption{(Color online) Particle-hole energy gap defined in
    Eq.~(\ref{coopergap}) as a function of the particle density for
    $\alpha/\pi=1/3$ (left panel), $\alpha/\pi=2/5$ (center panel), and
    $\alpha/\pi=1/2$ (right panel).  Depending on the value of $\alpha$,
    different peaks are seen in the energy gap that are symmetric around
    half-filling. Apart from the dominant peak at half-filling, $n=1/2$, we
    observe additional peaks at $n=1/6$ and $5/6$ for $\alpha/\pi=1/3$, $n=1/5$
    and $4/5$ for $\alpha/\pi=2/5$, and $n=1/4$ and $3/4$ for
    $\alpha/\pi=1/2$.}
  \label{dm}
\end{figure*}

Here, we limit DMRG calculations to hard-core bosons in the infinite-$U$
limit, providing an easier implementation of the algorithm as the Hilbert
space is drastically reduced by excluding multiple occupation of each site.
This limit is particularly important for correlated states, as the gaps in the
spectrum are expected to be more prominent with strong interactions.  Within
this constraint, each site is allowed to be empty or have only one boson so
that the maximum occupation number $n_{max}=1$ and the terms with the on-site
interaction in the Hamiltonian become only a constraint in the Hilbert space.
The Bose-Hubbard model in this limit can be mapped to a spin-XXZ model, where
the ground
state is at half-filling.  We find that our system has a ground state at half-filling not only for $\alpha=0$ but also for nonzero $\alpha$. In the two
limits, all sites empty and all sites filled, the ground-state energy is
0 and the minimum of the energy is always at half-filling, which is in the
middle of these two limits.

\begin{figure}
  \includegraphics[scale=0.8]{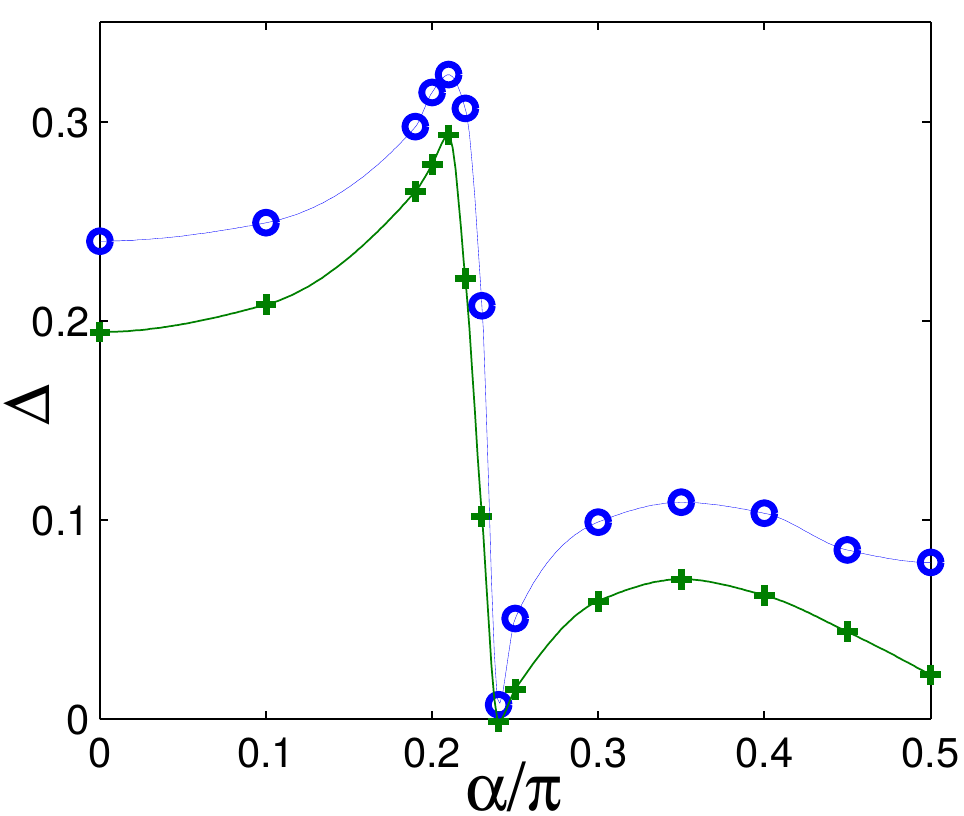} 
  \caption{(Color online) Gap between the ground state and two excited states
    as a function of the magnetic field at half-filling for hard-core
    interaction. The gap between the first excited state and the ground state
    ($E^1-E^0$) is shown by (green) crosses, whereas the gap between the
    second excited state and the ground state ($E^2-E^0$) is shown by (blue)
    circles.  Thin solid lines are spline interpolations to data points. The
    spectrum shows a jump at $\alpha_c/\pi=0.21$, which is very close to the
    critical magnetic field calculated from the single-particle spectrum.} 
  \label{dmrggap1}
\end{figure}

The energy gap between the ground-state and the first two excited states is shown
in Fig.~\ref{dmrggap1} for half-filling. The figure shows that the
spectrum of the three lowest lying states changes abruptly at
$\alpha_c/\pi\approx 0.21$. This plot is symmetric around $\alpha/\pi=0.5$ so
we only display the half. The critical value found here is consistent with the
one found in the single-particle solution, which is equal to 0.2148 or 0.7852. 

To get the energies at different fillings the DMRG code must be restricted to a
different particle-number-conserving subspace. We use the route proposed by
Ramanan \textit{et al.} \cite{pai3}, in which the plateaus in the chemical
potential versus the density plots and the corresponding compressibility are
obtained successfully.  We again have a system length $L=60$ and $2\times
L=120$ sites. Beginning from $L=4$ and a total number of particles $N=4$, we
increase both the lattice length and the number of particles up to where the
number of particles is $N=10$. After that, the lattice length is
increased while the total number of particles held fixed at 10. Whenever the
lattice length reaches $L=60$, finite system sweeps are used to decrease the
energy. Next, we increase the total particle number by 1, keeping  the system
size fixed, and perform five sweeps to get the energy for this new filling.
Repeating this procedure, we get energies where the particle number is
increased up to $N=110$. In the end, energies of systems from $N=10$ to
$N=110$ particles placed at $2\times L=120$ sites are obtained. After that,
the gap formula defined by Cooper \textit{et al.} \cite{coopervortex},
\begin{equation}
 \Delta=N\left[ \frac{E(N+1)}{N+1}+\frac{E(N-1)}{N-1}-2\frac{E(N)}{N}\right] 
\label{coopergap}
\end{equation}
is used, which minimizes finite-size effects. 

We show this gap for various values of magnetic field in Fig.~\ref{dm}. It is
shown that the gap oscillates between zero and nonzero values for low densities
and becomes negative towards integer filling. Apart from that there are three
dominant peaks; one is always at 1/2 and the other two depend on
$\alpha$. The magnitudes of these changing peaks are also shown to get
smaller and smaller as the field approaches 1/2. It is interesting to
compare these peaks by defining the filling factor \cite{rev_cooper},
\begin{equation}
 \nu=\frac{n}{f}
\label{filling}
\end{equation}
where $n$ is the particle density and $f$ is the vortex density defined as the phase
attained around a unit cell divided by $2\pi$, which means $f=\alpha/\pi$ in
our model.  We see that the corresponding distinct values of the filling
factors for the peaks in Fig.\ref{dm} are obtained as $\nu=1/4$, $3/4$, $5/4$
for $\alpha/\pi=1/3$; $\nu=1/4$, $5/8$, $1$ for $\alpha/\pi=2/5$; and
$\nu=1/4$, $1/2$, $3/4$ for $\alpha/\pi=1/2$.

The dependence of the gap on the filling fraction is clear evidence of the
role played by the interactions. However, our simple finite-size DMRG
calculations can not reveal the character of correlations within these states.
Future studies of the system in this limit must include larger system sizes,
finite on-site interactions, and a careful consideration of finite size
effects to reveal the physics of possible correlated states in the two-leg
Bose-Hubbard ladder.

\section{Conclusion}
\label{sec:concl}

Our calculations lead to a number of conclusions related to the recent
experiment in Ref.~\cite{THEEXPERIMENT}. 

The experiment probed only the limit where the number of particles per site is
high, which can mostly be described by the Gross-Pitaevskii level
approximations. The reported phase transition between the two phases  is
driven by the change in the character of the single-particle spectrum rather
than interactions. In this limit, the effect of interactions is expected to be
quantitative rather than qualitative. Our calculations indicate that the
interactions will shift the boundary between the Meissner and the vortex
phases, however, observation of this shift is complicated by the uncertainty
due to the finite temperature in the experiments. A recent paper
\cite{ErichTwoLeg} argues that another effect of the interactions would be the
spontaneous breaking of the symmetry between the two-legs. As our calculations
have this symmetry built in we cannot investigate such a transition.

While the current experiment operates in the superfluid regime, it is natural
to expect further experiments in this system to probe the region with only a
few particles per site where the insulating state is likely. We expect our
strong-coupling and DMRG results to be quantitatively  correct for the Mott
transition boundary.  While the effect of the external confining potential is
weak in the experiment, a wedding-cake structure would be a clear indication
of the Mott transition. Such a wedding-cake structure not only can be observed
by looking at the density but also can be deduced from the link currents
investigated by the method used in the present experiment.

Finally, our DMRG results for noninteger filling factors provide some
evidence for the possibility of correlated phases in this system. However, we
can not confidently assert the presence of these phases due to the finite-size
limitations of our calculation. To judge the viability of the experimental
observation of these phases a better characterization of their gaps and
correlation properties must be made. Nonetheless, our results indicate that
this regime should be interesting to investigate experimentally. 

In conclusion, we have worked on the two-leg Bose-Hubbard ladder exposed to a
magnetic field within various theoretical approaches and implemented  the DMRG
to study the behavior of the system. We have found that the system has two
distinctively different regimes, in agreement with the recent experiment. The
shape of the Mott insulator region is obtained by three methods: variational
mean-field theory, strong-coupling perturbation theory, and DMRG.  We found
that the shape of the lobe is consistent within the DMRG and the
strong-coupling approximation, while the results of the mean-field theory are
relatively poor. Apart from the determination of the Mott lobes, the system is
found to display novel physical properties as a result of the single-particle
spectrum. We believe that this model serves as an important tool for
understanding the general properties of optical lattices coupled to a gauge
field. In the latter part of the paper, we have calculated the excitation gap
for non-integer filling and found distinct peaks at simple fractions of
particle number to flux quanta. This regime will be investigated further in
subsequent work.

\textit{Note Added in Proof.}
Recently, the same system was
investigated theoretically in \cite{piraud2014vortex} and
\cite{PhysRevA.89.023619}. We believe that our results and these theoretical papers
are complementary.
\begin{acknowledgments}
A.K. thanks Levent Suba\c{s}{\i} and Onur Umucal{\i}lar for many
useful discussions, and Tomotoshi Nishino and Andrej Gendiar for useful
correspondence on DMRG. M.\"O.O. was supported by TUBITAK Grant No.  112T974. 
\end{acknowledgments}

\end{document}